\documentclass[journal=jacsat,manuscript=article]{achemso}

\usepackage{chemformula} 
\usepackage[T1]{fontenc} 
\usepackage{bm}
\usepackage{hyperref}
\usepackage[normalem]{ulem}

\author{Fen Zou}
\affiliation{Center for Theoretical Physics \& School of Physics and Optoelectronic Engineering, Hainan University, Haikou 570228, China}

\author{Yong Li}\email{yongli@hainanu.edu.cn}
\affiliation{Center for Theoretical Physics \& School of Physics and Optoelectronic Engineering, Hainan University, Haikou 570228, China}

\author{Peng Zhang}\email{pengzhang@ruc.edu.cn}
\affiliation[Unknown University]{Department of Physics, Renmin University of China, Beijing, 100872, China}
\alsoaffiliation[Second University]{Key Laboratory of Quantum State Construction and Manipulation (Ministry of Education), Renmin University of China, Beijing, 100872, China}

\title[An \textsf{achemso} demo]
{Enantiomer-Specific Pumping of Chiral Molecules}

\abbreviations{IR,NMR,UV}

\begin{document}

%
%
%
%
%

\begin{abstract}
Enantiomer-specific state transfer (ESST), which involves transferring enantiomers with different handedness of a chiral molecule into different-energy internal states, is a challenging yet significant task. Previous ESST methods are based on dynamic processes and thus require the preparation of initial states and precise control of microwave operation times. We propose a novel ESST approach, called enantiomer-specific pumping (ESP), which is based on a {\it dissipative} process, and thereby eliminates the need for these two technical requirements. This approach utilizes a special microwave-induced dark state that appears exclusively for the enantiomer with a specific handedness. Specifically, in ESP, the enantiomer lacking the dark state can be pumped out of the subspace of relevant internal states, while the enantiomer with the dark state maintains a finite probability within this subspace, offering high efficiency in ESST. Notably, ESP facilitates enantiodetection without the need for enantiopure samples as reference.
\end{abstract}

Chiral molecules contain a pair of enantiomers that are mirror images of each other, {\it i.e.}, molecules with left and right handedness. These two enantiomers have almost identical physical properties, but very different chemical properties and biological activities~\cite{mezey1991global}. Transferring different enantiomers of a kind of chiral molecule into different-energy internal states (enantiomer-specific state transfer, ESST) is a challenging task, and has attracted many attentions and efforts~\cite{Kral2001Cyclic,Li2008Dynamic,Jia2010Distinguishing,Leibscher2019Principles,Vitanov2019Highly,Ye2019Effective,Torosov2020Chiral,
Wu2020Two,Guo2022Cyclic,leibscher2022Fulla,Liu2022Enantiospecific,Cheng2023Enantiomer,Eibenberger2017Enantiomer,Perez2017Coherent,Perez2018State,Lee2022Quantitative,sun2023Inducing}. ESST can be used not only for studies of chiral-molecule properties~\cite{lovas2009Microwavea,milner2019Controlled}, but also for the detection of enantiomeric excess (enantiodetection of chiral molecules)~\cite{berova2000circular,nafie2011vibrational,stephens1985theory,barron2009molecular,Begzjav2019Enhanced,Jia2011Probing,Patterson2013Enantiomer,Patterson2013Sensitive,Patterson2014New,
Shubert2014Identifying,shubert2015Rotational,Lobsiger2015Molecular,Shubert2016Chiral,Yachmenev2016Detecting,Ye2019Determination,
Xu2020Enantiomeric,Ye2021Entanglement,cai2022Enantiodetection,Chen2022Enantiodetection,Kang2023Near,ye2023SingleShota} and the spatial separation of different enantiomers~\cite{Li2007Generalized,Li2010Theory,Eilam2013Spatial,Liu2021Spatial}.
In recent years, the ESST of chiral molecules has been demonstrated in several experiments~\cite{Eibenberger2017Enantiomer,Perez2017Coherent,Perez2018State,Lee2022Quantitative,sun2023Inducing}.

Current ESST approaches~\cite{Kral2001Cyclic,Li2008Dynamic,Jia2010Distinguishing,Leibscher2019Principles,Vitanov2019Highly,Ye2019Effective,Torosov2020Chiral,
Wu2020Two,Guo2022Cyclic,leibscher2022Fulla,Liu2022Enantiospecific,Cheng2023Enantiomer,Eibenberger2017Enantiomer,Perez2017Coherent,Perez2018State,Lee2022Quantitative,sun2023Inducing} are based on unitary evolution of molecular internal states, commonly via cyclic three-level structures of chiral molecules. As a result, to apply these approaches one requires to prepare initially the molecules in a certain internal state of the three-level structure. Additionally, in most of these approaches the operation times of the laser or microwaves should be controlled precisely. The initial-state preparation and operation-time control can introduce complexities and errors into the experiments of ESST.

In this Letter we propose an ESST approach for chiral molecules in gas phase, with the initial-state preparation and operation-time control being {\it not required}. Explicitly, in our approach the enantiomers with different handedness are transferred to different-energy states via a {\it dissipative} evolution, rather than a unitary evolution as above. We call this approach as enantiomer-specific pumping (ESP).

\begin{figure}[t]
	\centering
	\includegraphics[width=12cm]{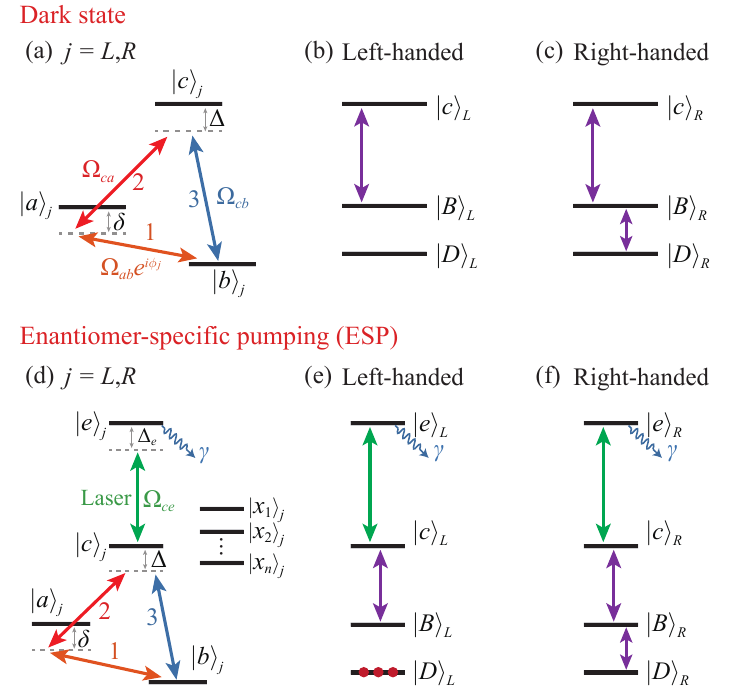}
        \caption{
        {\bf (a):} The three-level structure of chiral molecules with cyclic electric-dipole transitions.
        {\bf (b):} $|D\rangle_{L}$ is not coupled to $|B\rangle_L$ and $|c\rangle_L$, and thus is a dark state of the left-handed molecule.
        {\bf (c):} $|D\rangle_{R}$ is coupled to $|B\rangle_R$, and $|B\rangle_R$ is coupled to $|c\rangle_R$. Thus, $|D\rangle_{R}$ is not a dark state of the right-handed molecule.
        {\bf (d):} Energy levels of ESP.
        {\bf (e, f):} A qualitative explanation for the principle of ESP.
        For details, see the main text.
        }
	\label{fig1}
\end{figure}

As most of current ESST approaches, our scheme is built upon a cyclic three-level structure with internal states $a$, $b$, and $c$ being coupled with each other by three microwaves [Fig.~\ref{fig1}(a)], where the overall phases of the three Rabi frequencies for two enantiomers differ by $\pi$~\cite{Kral2001Cyclic}.
We show that due to this $\pi$ phase difference, when the microwaves satisfy appropriate conditions, the molecule with a certain handedness has an eigen-state (dark state), which is a superposition of $a$ and $b$ states and does not include the component of the $c$ state, meanwhile the molecule with the other handedness has not such a dark state.  When a laser beam is further applied to couple the $c$ state to an upper level $e$ in the electronically-excited potential surface [Fig.~\ref{fig1}(d)], due to the laser/microwave induced couplings and the spontaneous decay of the $e$ state, the enantiomer without the dark state is pumped out from the $a$ and $b$ states, while the other one with the dark state still has finite probability in these two states. In another word, if one molecule is found to be in $a$ or $b$ states, it definitely has the handedness with the dark state. That is the principle of ESP.

With the help of ESP one can increase the efficiency of ESST. The ESP can be further applied for enantiodetection of chiral molecules and spatial separation of different enantiomers. In particular, using ESP one can realize enantiodetection, even without enantiopure samples as reference.

\textit{Dark state.}
Let us consider a left- (right-) handed three-level chiral molecule $L$ ($R$), as shown in Fig.~\ref{fig1}(a). 
We consider three rovibrational states $|a\rangle_j$, $|b\rangle_j$ and $|c\rangle_j$ 
for the molecule $j$ ($j=L, R$) ({\it e.g.}, $|a,b,c\rangle_j$ are three rotational states of the lowest vibrational level). The wave function of the nuclei and electrons of the state $|s\rangle_L$ is 
the spatial inversion of the one of $|s\rangle_R$ $(s=a,b,c)$ (up to a global phase factor). Moreover, three microwave beams~\cite{3WV} are coupled to both the molecules $L$ and $R$. Explicitly, the beams 1, 2, and 3 are near-resonant to the electric-dipole
transitions $a \leftrightarrow b$, $a \leftrightarrow c$, and $b \leftrightarrow c$, respectively. Under the rotating-wave approximation, the original Hamiltonian ${\hat H}^{(j)}_{\text{ori}}$ of the molecule $L$ ($R$) is given by ($\hbar=1$):
\begin{eqnarray}
{\hat H}^{(j)}_{\text{ori}}&=&\sum_{s=a,b,c}E_s|s\rangle_{j}\langle s|+\frac{1}{2}\bigg[
{\Omega}_{ab}e^{i\phi_{j}}|a\rangle_{j}\langle b|e^{-i\omega_1 t}
\bigg.\nonumber\\
&&\bigg.+{\Omega}_{ca}|c\rangle_{j}\langle a|e^{-i\omega_2 t}
+{\Omega}_{cb}|c\rangle_{j}\langle b|e^{-i\omega_3 t}+\text{H.c.}\bigg],\hspace{1.5cm} (j=L,R),\label{h}
\end{eqnarray}
where $E_s$ ($s=a,b,c$) and $\omega_\alpha$ ($\alpha=1,2,3$) are the energy of the molecular states $|s\rangle_{L,R}$ and the angular frequencies of the microwave beams $\alpha$, respectively. Without loss of generality, we assume $E_{c}>E_a>E_b$. Additionally, in Eq.~(\ref{h}), $\Omega_{ab}e^{i\phi_{L(R)}}$, $\Omega_{ca}$, and $\Omega_{cb}$ are the strengths (Rabi frequencies) of the electric-dipole couplings induced by the microwaves 1, 2, and 3, respectively, for molecule $L$ ($R$). Due to the spatial-inversion relation between the states $|s\rangle_{L}$ and $|s\rangle_{R}$ ($s=a,b,c$), without loss of generality, we assume the amplitudes $\Omega_{ab}$, $\Omega_{ca}$, and $\Omega_{cb}$ are all {\it real and positive}, and are the same for both the molecules $L$ and $R$, while the handedness-dependent overall phases $\phi_{L,R}$ satisfy $\phi_{R}=\phi_{L}+\pi$ ~\cite{supp}. Under this gauge, the overall phase $\phi_{L}$, the angular frequencies $\omega_{1,2,3}$, and the amplitudes $\Omega_{ab}$, $\Omega_{ca}$, and $\Omega_{cb}$ are independent control parameters of our system.

We assume the overall phase $\phi_{L}$ of the microwave beams satisfies:
\begin{eqnarray}
\phi_{L}&=&0,\label{con0}
\end{eqnarray}
and the angular frequencies $\omega_{1,2,3}$ satisfy the three-photon resonant condition, i.e.,
\begin{eqnarray}
\omega_{3}&=&\omega_{1}+\omega_2.\label{con1}
\end{eqnarray}
Under these conditions, in the interaction picture with respect to ${\hat H}^{(j)}_{0}=E_{b}|b\rangle_{j}\langle b|+(E_{b}+\omega_{1})|a\rangle_{j}\langle a|+(E_{b}+\omega_{3})|c\rangle_{j}\langle c|$ the Hamiltonian of this system becomes time-independent:
\begin{eqnarray}
{\hat H}^{(j)}&=& \Delta |c\rangle_{j}\langle c|+\delta|a\rangle_{j}\langle a|
\nonumber\\
&&+\frac {1}{2}\bigg[
{\Omega}_{ca}|c\rangle_{j}\langle a|\!+\!{\Omega}_{cb}|c\rangle_{j}\langle b|\!+\!\xi^{(j)} {\Omega}_{ab}|a\rangle_{j}\langle b|\!+\!\text{H.c.}\bigg],\nonumber\\
&&\hspace{5cm} (j=L,R), \label{hi}
\end{eqnarray}
where $\Delta=E_c-E_b-\omega_3$, $\delta=E_a-E_b-\omega_1$, and $\xi^{(L/R)}=\pm 1$.

It is convenience to perform the following discussion in the basis of $\{|c\rangle_{j}, |D\rangle_{j}, |B\rangle_{j}\}$, with the states $|B,D\rangle_{j}$ being defined as
\begin{eqnarray}
|D\rangle_j&\equiv& {\cal Z}^{-1}\big[\Omega_{cb}|a\rangle_j-\Omega_{ca}|b\rangle_j\big], \label{D}\\
|B\rangle_j&\equiv& {\cal Z}^{-1}\big[\Omega_{ca}|a\rangle_j+\Omega_{cb}|b\rangle_j\big], \ \ (j=L,R),\label{B}
\end{eqnarray}
where ${\cal Z}={\sqrt{\Omega_{cb}^2+\Omega_{ca}^2}}$. With straightforward calculations we find that
$_j\langle c|{\hat H}^{(j)}|B\rangle_j={\cal Z}/2>0$ and $_j\langle c|{\hat H}^{(j)}|D\rangle_j=0$ ($j=L,R)$, {\it i.e.}, the microwaves directly couple
the state $|c\rangle_{L(R)}$ only to $|B\rangle_{L(R)}$, and not  to $|D\rangle_{L(R)}$.
Additionally, the microwave 1 also induces the coupling between $|D\rangle_{L(R)}$ and $|B\rangle_{L(R)}$, with coupling strength:
\begin{eqnarray}
_j\langle B|{\hat H}^{(j)}|D\rangle_j&=&\frac{1}{{\cal Z}^2}\bigg[
\delta\Omega_{cb}\Omega_{ca}+\frac{1}{2}\xi^{(j)}\Omega_{ab}\left(\Omega_{cb}^2-\Omega_{ca}^2\right)
\bigg],\nonumber\\
&&\hspace{3.5cm}(j=L,R).\label{cs}
\end{eqnarray}
Notice that this coupling strength is handedness-dependent through the factor $\xi^{(j)}$. Under the condition
\begin{eqnarray}
\delta=\frac{\Omega_{ab}\left(\Omega_{ca}^2-\Omega_{cb}^2\right)}{2\Omega_{cb}\Omega_{ca}}=:\delta_{0}
\neq 0,\label{con4} 
\end{eqnarray}
we have $_L\langle B|{\hat H}^{(L)}|D\rangle_L=0$, but $_R\langle B|{\hat H}^{(R)}|D\rangle_R\neq 0$. 
Since we also have
$_L\!\langle c|{\hat H}^{(L)}|D\rangle_L=0$, in this case
the state $|D\rangle_L$ is decoupled to both $|c\rangle_L$ and $|B\rangle_L$ [Fig.~\ref{fig1}(b)], and thus is an eigen-state of ${\hat H}^{(L)}$. However, the state $|D\rangle_R$ is still coupled to $|B\rangle_R$ [Fig.~\ref{fig1}(c)].

The state $|D\rangle_L$, which is an eigen-state of ${\hat H}^{(L)}$, is a ``dark state'', because it is a superposition of $|a\rangle_L$ and $|b\rangle_L$, is not coupled to $|c\rangle_L$ by the microwaves.
Furthermore, this dark state appears exclusively for the left-handed molecule $L$, as the Hamiltonian ${\hat H}^{(R)}$ of the right-handed molecule does not possess such an eigenstate.
Moreover, under the conditions~(\ref{con1}, \ref{con4}), and $\phi_{L}=\pi$, a dark state appears only for the right-handed molecule. When $\phi_{L}=\pi$ ($\phi_{L}=0$), a dark state can appear only for the left- (right-) handed molecule, under the appropriate conditions, which can be derived straightforwardly via the above approach.

\textit{Enantiomer-specific pumping.}
Now we introduce the enantiomer-specific pumping (ESP), which occurs when an appropriate laser beam is applied for the above system under the conditions of Eqs.~(\ref{con0}, \ref{con1}, \ref{con4}). This laser beam couples the state $|c\rangle_{L(R)}$ to an additional state $|e\rangle_{L(R)}$  on the electronically-excited potential surface [Fig.~\ref{fig1}(d)]. Via spontaneous radiation, the molecule in state $|e\rangle_{L(R)}$ can decay back to the states on the electronically-ground potential surface, including the states $|a,b,c\rangle_{L(R)}$
and other vibrational and rotational levels which are called as $|x_1\rangle_{L(R)},...,|x_n\rangle_{L(R)}$ [Fig.~\ref{fig1}(d)].

For the right-handed molecule $R$ [Fig.~\ref{fig1}(f)],  as mentioned above, the microwaves can induce both the  $|c\rangle_R\leftrightarrow|B\rangle_R$ and $|B\rangle_R\leftrightarrow|D\rangle_R$ couplings. Additionally, the laser beam induces the $|c\rangle_R\leftrightarrow|e\rangle_R$ coupling. As a result, when the molecule is in the states $|a\rangle_R$ or $|b\rangle_R$, it would transit to $|e\rangle_R$ via these couplings, and then decay to all the electronic ground states $|a,b,c\rangle_{R}$ and $|x_{1,...,n}\rangle_{R}$. This process can be repeated many times, and
once the molecule transits to $|x_{1,...,n}\rangle_{R}$, it cannot return to $|a,b\rangle_{R}$~\cite{note}. Finally, the population probabilities of both the states $|a\rangle_R$ and $|b\rangle_R$ tend to zero.

However, the situation for the left-handed molecule $L$ is quite different [Fig.~\ref{fig1}(e)]. Since the dark state $|D\rangle_L$ is not coupled to the states $|B\rangle_L$ and $|c\rangle_L$, the molecule in $|D\rangle_L$ cannot transit to the excited state $|e\rangle_L$, even when the laser is applied. Due to this fact, provided that initially the molecule $L$ has a non-zero probability on the dark state $|D\rangle_L$, the population probabilities of this molecule in the $|a\rangle_R$ and $|b\rangle_R$ states are always non-zero, even after a long evolution time.

In summary, when the laser is applied, the right-handed molecule is completely pumped out from the states $a$ and $b$, while the left-handed molecule cannot be totally pumped out from these two states due to the existence of the dark state. Thus, one achieves the ESST via the above ESP scheme by focusing on the populations of the $a$ and $b$ states, regardless the initial states and precise operation time.

In realistic molecular gases, there are inter-molecule collisions, which can induce the internal-state transitions and decoherence effects for the molecules. In particular, when the molecule is in the states $|x_{1,...,n}\rangle_{L(R)}$, it can transit back to the states $|a,b\rangle_{L(R)}$ via inelastic collisions. As a result, even in the long-time limit the right-handed molecule still has finite probability to be in the $a$ and $b$ states. However, when the microwaves, the laser, and the spontaneous decay of the state $|e\rangle_{L(R)}$ are strong enough, the transitions induced by the laser and microwaves and the spontaneous radiation can be much faster than the collisional transitions. In this case, our above results of ESP are not qualitatively changed. {I.e.}, for a long-enough evolution time, the populations of the left-handed molecules being in the states $a$ and $b$ are much larger than the ones of the right-handed molecules.

\textit{Illustration of ESP.}
We further illustrate the effect of ESP via numerically solving the quantum master equation~\cite{supp} for the density matrix $\hat{\rho}^{(j)}$ $(j=L,R)$ of molecule $j$ in a molecular gas. We focus on the case with $\Delta=0$, where the master equation is given by
\begin{equation}\label{QME}
\frac{d\hat{\rho}^{(j)}}{dt}=-i[\hat{H}^{\prime(j)},\hat{\rho}^{(j)}]
+\mathcal{L}[\hat{\rho}^{(j)}]+\mathcal{D}[\hat{\rho}^{(j)}].
\end{equation}
Here $\hat{H}^{\prime(j)}=\hat{H}^{(j)}+\Delta_{e}|e\rangle_{j}\langle e|+\Omega_{ce}(|c\rangle_{j}\langle e|+|e\rangle_{j}\langle c|)/2$, with $\Omega_{ce}$ being the Rabi frequency of the laser-induced $c-e$ coupling, and $\Delta_{e}=E_e-E_c-\omega_4$, where $E_{e}$ is the energy of the excited state $|e\rangle_{j}$, and $\omega_{4}$ is the angular frequency of the laser beam.
Additionally,
$\mathcal{L}[\hat{\rho}^{(j)}]$ and $\mathcal{D}[\hat{\rho}^{(j)}]$ describe the effects from the spontaneous decay of  $|e\rangle_j$ and the inter-molecule collisions, respectively. There are various channels of the spontaneous decay and collision, corresponding to different finial states.
The branch ratios of these channels are determined by the details of the molecular structure. Here, we make the unbiased assumption that the ratios of all decay (collision) channels are equal. As a result,
$\mathcal{L}[\hat{\rho}^{(j)}]=\gamma(\sum_{s=c,a,b}\mathcal{L}_{|s\rangle_{j}\langle e|}[\hat{\rho}^{(j)}]+\sum_{k=1}^{n}\mathcal{L}_{|x_{k}\rangle_{j}\langle e|}[\hat{\rho}^{(j)}])/(n+3)$~\cite{note}, with $\gamma$ being the total decay rate of $|e\rangle_j$, and $\mathcal{L}_{\hat{o}}[\hat{\rho}^{(j)}]\equiv [2\hat{o}\hat{\rho}^{(j)}\hat{o}^{\dagger}
-\hat{o}^{\dagger}\hat{o}\hat{\rho}^{(j)}-\hat{\rho}^{(j)}\hat{o}^{\dagger}\hat{o}]/2$,
and  $\mathcal{D}[\hat{\rho}^{(j)}]=\kappa[{\hat I}/(n+4)-\hat{\rho}^{(j)}]$ \cite{supp}, with ${\hat I}$ being the identical operator, and
$\kappa$ being the collisional relaxation rate.

\begin{figure}[t]
 \centering
 \includegraphics[width=16cm]{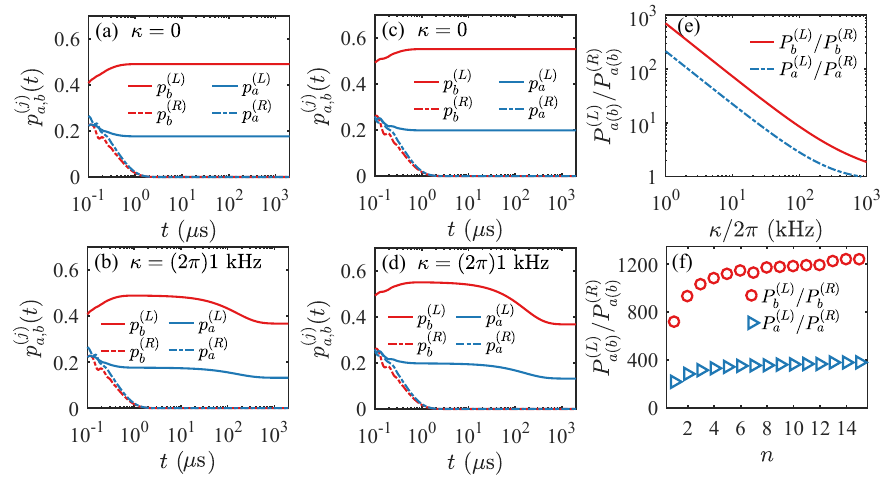}
        \caption{
        {\bf (a-d):} The populations $p_{a,b}^{(j)}(t)$ ($j=L,R$) of the molecule $j$ in states $a,b$, for the cases with the amount $n$ of the $x$-state being $n=1$, the collisional relaxation rate being $\kappa=0$ (a, c) or $\kappa=(2\pi)$\,1\,kHz (b, d), and the initial state ${\hat \rho}^{(j)}(0)=(|a\rangle_j\langle a|+|b\rangle_j\langle b|+|c\rangle_j\langle c|)/3$ (a, b) or ${\hat \rho}^{(j)}(0)=(|a\rangle_j\langle a|+|b\rangle_j\langle b|)/2$ (c, d).
        {\bf (e):} The  steady-state population ratio $P_{a(b)}^{(L)}/P_{a(b)}^{(R)}$ as functions of $\kappa$, for  $n=1$.
        {\bf (f):} The ratios $P_{a(b)}^{(L)}/P_{a(b)}^{(R)}$ as functions of $n$, for $\kappa=(2\pi)$\,1\,kHz. Other parameters are:
        $\phi_{L}=0$, $\delta=\delta_{0}$, $\Delta=\Delta_{e}=0$, $\Omega_{ca}=\Omega_{ab}=(2\pi)\,10\,\text{MHz}$, $\Omega_{cb}=(2\pi)\,6\,\text{MHz}$, $\Omega_{ce}=(2\pi)\,20\,\text{MHz}$, and $\gamma=(2\pi)\,10\,\text{MHz}$.
        }
 \label{fig2}
\end{figure}

By solving Eq.~(\ref{QME}), we calculate the time-dependent population $p_{s}^{(j)}(t)=\,_{j}\langle s|\hat{\rho}^{(j)}(t)\vert s\rangle_{j}$ of the internal state $s$ ($s=a,b$) for molecule $j$ ($j=L, R$).
In Figs.~\ref{fig2}(a-d) we show  $p_{a,b}^{(L,R)}(t)$
for typical cases with various initial state $\hat \rho^{(j)}(0)$ and collisional relaxation rate $\kappa$,
where the dark state appears for the molecule $L$, and the decay rate $\gamma$ and the Rabi frequencies are of the order of $(2\pi)\,10\,\text{MHz}$ (with explicit values being given in the caption). It is shown that in each case the populations $p_{a,b}^{(R)}(t)$ of the molecule $R$ decay to almost zero within a very short evolution time ($\sim 1\,\mu$s). This result can be explained as follows. As introduced above, the right-handed molecules are pumped out from the states $a$ and $b$ via the laser/microwave induced transitions and the spontaneous decay of the $e$ state. As a result, the time required for the system to reach a significant ESP effect is of the order of the characteristic time of these processes. Notice that this time scale is much less than the time required for the molecules to relax to the steady state ($\sim 1/\kappa$)~\cite{ke0}. In the supplementary document~\cite{supp}, we further illustrate $p_{a,b}^{(L,R)}(t)$ for two other initial states, with the behaviors being similar as the ones of Figs.~\ref{fig2}(a-d), and illustrate the time evolution of the populations $p_{D,B}^{(j)}(t)$ ($j=L,R$) of  the dressed states $|D,B\rangle_{j}$ for the cases of Figs.~\ref{fig2}(c, d). It is shown that the populations $p_{D,B}^{(R)}(t)$ and $p_{B}^{(L)}(t)$ decay to nearly zero within a very short evolution time (approximately $1\,\mu$s), while the steady-state population $p_{D}^{(L)}$ remains non-zero. This is consistent with the above physical picture of the ESP.

We further derive the steady-state population $P_{s}^{(j)}\equiv p_{s}^{(j)}(t\rightarrow\infty)$ of the state $s$ ($s=a,b$) for molecule $j$ ($j=L, R$), and illustrate the ratios $P_{a(b)}^{(L)}/P_{a(b)}^{(R)}$ as functions of $\kappa$ in Fig.~\ref{fig2}(e), for the cases with other parameters being the same as those in Figs.~\ref{fig2}(a-d). It is clearly shown that these ratios are over $10^2$ for $\kappa=(2\pi)\,1\,\text{kHz}$, and decrease with the collisional relaxation rate $\kappa$. Thus, to realize ESP in a realistic molecular gas, one requires to tune the collision rate to be small enough by changing, {\it e.g.}, the molecular density and temperature.
Since the results of Fig.~\ref{fig2}(e) are for the laser detuning $\Delta_e$ being zero, in the Supplementary document~\cite{supp} we further illustrate $P_{a(b)}^{(L)}/P_{a(b)}^{(R)}$ as functions of $\kappa$, for $\Delta_e=\pm(2\pi)\,25\,\text{MHz}$ and $\pm(2\pi)\,50\,\text{MHz}$. It is shown that for these cases, when $\kappa\leq(2\pi)\,1\,\text{kHz}$, the ESP effect is also significant, with $P_{a(b)}^{(L)}/P_{a(b)}^{(R)}\geq 10$.

The above results of Figs.~\ref{fig2}(a-e) are for the case with a single $x$-state ($n=1$). In Fig.~\ref{fig2}(f) we further show the ratio of steady-state populations $P_{a(b)}^{(L)}/P_{a(b)}^{(R)}$ for the cases with  $1\leq n\leq 15$, and other parameters being the same as those in Fig.~\ref{fig2}(b). It is shown that $P_{a(b)}^{(L)}/P_{a(b)}^{(R)}$ slightly increases with $n$. This implies that the ESP effect is qualitatively robust with respect to the amount of the states to which the molecules can decay from the $e$ state.

\textit{Applications of ESP.}
The effect of ESP is independent of the molecular initial state and robust for the operation times of the laser and microwaves. For instance, in the systems of Figs.~\ref{fig2}(a-d), there is significant ESP effect for each initial state, when the evolution time is of the order of $\mu$s or longer.
As a result, the initial-state preparation and the precise control of operation time are not required, and thus the corresponding errors are waved. In the examples of Figs.~\ref{fig2}(b, d), the steady-state purities  of left handedness in internal states $a$ and $b$, which are defined as
$\varepsilon_{s}\equiv P_{s}^{(L)}/(P_{s}^{(L)}+P_{s}^{(R)})$ ($s=a,b$), are as high as $\varepsilon_{a}\approx \varepsilon_{b}\approx 99.99\%$.

Additionally, as mentioned before, for the systems of Fig.~\ref{fig2}, when the absolute value $|\Delta_e|$ of the laser detuning is increased from $0$ to $(2\pi)\,50\,\text{MHz}$, the ESP effect is still strong (see Fig.~S3 in the Supplementary document~\cite{supp}), although not as significant as the one for $\Delta_e=0$. This result yields that ESP has a degree of robustness to the Doppler broadening of the laser frequency.

\begin{figure}[tbp]
	\centering
	\includegraphics[width=10.4cm]{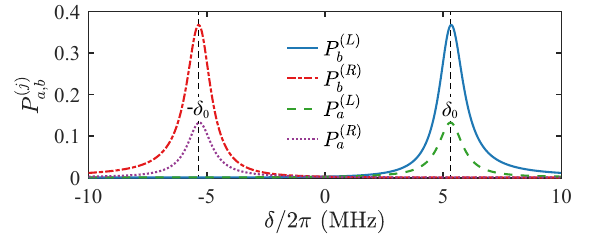}
        \caption{The steady-state populations $P_{a,b}^{(j)}$ ($j=L,R$) as functions of detuning $\delta$. Here we consider the cases with $n=1$ and $\kappa=(2\pi)\,1\,\text{kHz}$, and other parameters are the same as those in Fig.~\ref{fig2}.}
	\label{fig3}
\end{figure}

ESP can be further used to study chiral-molecule properties, spatial separation of different enantiomers~\cite{cs}, or enantiodetection of chiral molecules. In particular, for enantiodetection, one can apply the microwave and laser beams as above, and tune the detuning $\delta$ by changing the microwave frequencies $\omega_{1,2}$, with the conditions (\ref{con0}, \ref{con1}) always being satisfied and all the other parameters being fixed.
It is only required to measure the steady-state population $P_a(\delta)$ or $P_b(\delta)$ of the $a$ or $b$ state, as a function of $\delta$.
As shown in Fig.~\ref{fig3}, due to ESP, the steady-state populations $P_{s}^{(L)}$ ($P_{s}^{(R)}$) ($s=a, b$) of the left- (right-) handed molecules reach a maximum at $\delta=\delta_{0}$ ($\delta=-\delta_{0}$).
Thus, in a mixture of chiral molecules, the measured $P_{a(b)}(\delta)$ have two peaks, and the percentages of the molecules with each handedness can be derived from the ratio of the height of these two peaks. Notice that in this approach one does not need to know the exact values of $\delta_0$ before the detection. Thus, the details of the curves $P_{a,b}^{(L)}(\delta)$ [$P_{a,b}^{(R)}(\delta)$] of pure left- or right-handed molecules are not required for the enantiodetection.

\textit{Physical Realization.} ESP of chiral molecules is  experimentally feasible and needs to be performed in the gas phase.  Referring to recent experimental studies on quantitative study of ESST~\cite{Lee2022Quantitative}, we  choose 1-indanol as an example.  Explicitly, 
we choose the internal states $|a,b,c\rangle$ as
three rotational states in the lowest vibrational level:
$|b\rangle=|0_{0,0,0}\rangle$, $|a\rangle=|1_{0,1,0}\rangle$, and $|c\rangle=(|1_{1,0,1}\rangle+|1_{1,0,-1}\rangle)/\sqrt{2}$, respectively. Here  $|J_{K_{a},K_{c},M}\rangle$ denotes the rotational state with the angular momentum quantum number $J$ and the magnetic quantum number $M$. 
The  cyclic three-level structure with these three states has been realized and utilized  in the experiment~\cite{Lee2022Quantitative} of ESST.
Based on the rotational constants of 1-indanol $A=2\pi\times2410.071\,\textrm{MHz}$,  $B=2\pi\times1231.257\,\textrm{MHz}$, and $C=2\pi\times846.356\,\textrm{MHz}$~\cite{Lee2022Quantitative,HernandezCastillo2021}, the energy differences between the three internal states are $E_{c}-E_{b}=2\pi\times3641.328\,\textrm{MHz}$, $E_{a}-E_{b}=2\pi\times2077.613\,\textrm{MHz}$, and $E_{c}-E_{a}=2\pi\times1563.715\,\textrm{MHz}$.  To realize our ESP with these three states, one can apply three  microwaves which are linearly polarized along the $z$-, $y$-, and $x$-directions, respectively, to induce the transitions $|b\rangle\leftrightarrow|a\rangle$, $|b\rangle\leftrightarrow|c\rangle$, and $|a\rangle\leftrightarrow|c\rangle$, and apply a uv laser  to couple the state $|c\rangle$ to an electronic excited state, as in the experiment on quantitative study of ESST~\cite{Lee2022Quantitative}. Alternatively, we can choose states $|a,b,c\rangle$ as $|b\rangle=|0_{0,0,0}\rangle$, $|a\rangle=|1_{0,1,0}\rangle$, and $|c\rangle=|1_{1,0,1}\rangle$, respectively.  For these three states, one can apply a linearly $z$-polarized ($\sigma=0$) and two circularly polarized ($\sigma=1$) microwaves~\cite{Ye2018Real} to induce the transitions $|b\rangle\leftrightarrow|a\rangle$, $|b\rangle\leftrightarrow|c\rangle$, and $|a\rangle\leftrightarrow|c\rangle$, respectively.

\begin{acknowledgement}
The authors thank the Supporting by the Innovation Program for Quantum Science and Technology (Grant No.~2023ZD0300700), the National Key Research and Development Program of China (Grant No.~2022YFA1405300), the National Natural Science Foundation of China (Grants No.~12074030, No.~12274107, and No.~12405011) and the Research Funds of Hainan University [Grant No.~KYQD(ZR)23010].
\end{acknowledgement}

\begin{suppinfo}
Derivation of the collision term $\mathcal{D}[\hat{\rho}^{(j)}]$ of the quantum master equation (\ref{QME}), the populations $p_{D,B}^{(j)}(t)$ ($j=L,R$) of the molecule $j$ in states $|D,B\rangle_{j}$, the populations $p_{a,b}^{(j)}(t)$ ($j=L,R$) for various initial states, and the steady-state population ratios $P_{s}^{(L)}/P_{s}^{(R)}$ ($s=a,b$) for $\Delta_{e}\neq 0$.
\end{suppinfo}


\begin{mcitethebibliography}{53}
\bibitem[3WV()]{3WV}
	{W}e focus on microwave transitions because the phases of three microwave beams
	are easier to be locked. On the other hand, the frequencies of purely
	rotational transitions between molecular rotational states are typically on
	the order of microwave. Additionally, the required phase-locking can also be
	implemented by using a combination of different types of transitions, e.g.
	with applying two infrared lasers (coupling the rovibrational transitions)
	and one microwave beam (coupling the purely rotational transition).\relax
	\mciteBstWouldAddEndPunctfalse
	\mciteSetBstMidEndSepPunct{\mcitedefaultmidpunct}
	{}{\mcitedefaultseppunct}\relax
	\EndOfBibitem
\bibitem[not()]{note}
	{N}ote that here we have neglected the spontaneous decay among the states
	$|a,b,c\rangle$ and $|x_{1,...,n}\rangle$, because these decay rates are very
	small compared to the spontaneous decay rate of $|e\rangle$.\relax
	\mciteBstWouldAddEndPunctfalse
	\mciteSetBstMidEndSepPunct{\mcitedefaultmidpunct}
	{}{\mcitedefaultseppunct}\relax
	\EndOfBibitem
\bibitem[ke0()]{ke0}
	{F}or $\kappa=0$, the steady state $\hat \rho^{(j)} (t\rightarrow\infty)$
	($j={L,R}$) depends on the initial state, and the time scale of the system
	relaxing to this state is same as the one of $p_{a,b}^{(R)}(t)$ decaying to
	zero, as shown in {F}ig.~\ref{fig2}(a).\relax
	\mciteBstWouldAddEndPunctfalse
	\mciteSetBstMidEndSepPunct{\mcitedefaultmidpunct}
	{}{\mcitedefaultseppunct}\relax
	\EndOfBibitem
\bibitem[cs()]{cs}
	{A}fter achieving ESP (with a dark state for the left-handed molecules, for
	example) in a mixture of left-handed and right-handed molecules, it is
	possible to spatially separate the enantiomers via either of the following
	two approaches: (i) Separate the molecules in internal states $a$ or $b$ via
	methods such as laser ionization, laser dissociation, or
	internal-state-dependent chemical reactions. (ii) Transfer the left-handed
	molecules in states $a$ or $b$ to another internal state with a special
	electric dipole moment using a resonant laser or microwave beam, without
	affecting the right-handed molecules, and then spatially separate the
	left-handed molecules by allowing them to pass through a static electric
	field with a spatial gradient.\relax
	\mciteBstWouldAddEndPunctfalse
	\mciteSetBstMidEndSepPunct{\mcitedefaultmidpunct}
	{}{\mcitedefaultseppunct}\relax
	\EndOfBibitem
\end{mcitethebibliography}
\providecommand{\latin}[1]{#1}
\makeatletter
\providecommand{\doi}
{\begingroup\let\do\@makeother\dospecials
	\catcode`\{=1 \catcode`\}=2 \doi@aux}
\providecommand{\doi@aux}[1]{\endgroup\texttt{#1}}
\makeatother
\providecommand*\mcitethebibliography{\thebibliography}
\csname @ifundefined\endcsname{endmcitethebibliography}
{\let\endmcitethebibliography\endthebibliography}{}

\end{document}